\documentclass[preprint]{aastex62}

\usepackage{natbib}
\usepackage{amsmath}

\received{July 31, 2019}
\revised{September 2, 2019}
\accepted{September 3, 2019}

\submitjournal{ApJ}

\shorttitle{Spectropolarimetry Mg~{\sc ii} UV doublet}
\shortauthors{Manso Sainz et al.}

\begin{document}

\title{Spectropolarimetry of the Solar Mg~{\sc ii} h and k Lines}

\author{R. Manso Sainz}
\affil{Max-Planck-Institut f\"ur Sonnensystemforschung \\
Justuts-von-Liebig-Weg 3, 37077 \\
G\"ottingen, Germany}

\author{T. del Pino Alem\'an}
\affil{Instituto de Astrof\'\i sica de Canarias \\
Av. V\'\i a L\'actea s/n, 38205 \\
La Laguna, Tenerife, Spain}

\author{R. Casini}
\affil{High Altitude Observatory, National Center for Atmospheric Research
\\
PO-Box 3000, Boulder, CO 80307-3000, USA}

\author{S. McIntosh}
\affil{High Altitude Observatory, National Center for Atmospheric Research
\\
PO-Box 3000, Boulder, CO 80307-3000, USA}

\begin{abstract}

We report on spectropolarimetric observations across the Mg~{\sc ii}
h and k-lines at 2800\,\AA\ made by the Ultraviolet Spectrometer 
and Polarimeter onboard the Solar Maximum Mission satellite.
Our analysis confirms the strong linear polarization in the wings of both lines 
observed near the limb, as previously reported, but also demonstrates the presence 
of a negatively (i.e., radially oriented) polarized signal between the two lines. 
We find evidence for fluctuations of the polarization pattern over a broad spectral 
range, resulting in some depolarization with respect to the pure scattering case
when observed at very low spatial and temporal resolutions.
This is consistent with recent theoretical modeling that predicts this to be
the result of redistribution effects, quantum interference between the atomic 
levels of the upper term, and magneto-optical effects.
A first attempt at a quantitative exploitation of these signals 
for the diagnosis of magnetic fields in the chromosphere is attempted.
In active regions, we present observations of circular polarization 
dominated by the Zeeman effect.
We are able to constrain the magnetic field strength in the upper active chromosphere
using an analysis based on the magnetograph formula, 
as justified by theoretical modeling.
We inferred a significantly strong magnetic field ($\sim 500\,$G) 
at the 2.5\,$\sigma$ level on an exceptionally active, flaring region. 

\end{abstract}


\section{Introduction}

The Ultraviolet Spectrometer and Polarimeter 
\citep[UVSP;][]{Calvert+79, Woodgate+80}, on board the 
{\em Solar Maximum Mission} 
\citep[SMM;][]{Bohlin+80, Strong+99} provided the first 
successful opportunity to investigate the Sun through ultraviolet (UV) spectropolarimetry. 
%\citep{Mandelstam67}
Observations in the UV allow access to the highest 
and more energetic regions of the solar chromosphere and transition region (TR);
polarimetry of spectral lines in that spectral window offer the possibility
of probing the magnetic field and the highly anisotropic, dynamic processes
taking place in these regions.
%\citep{Tousey+73, Doschek+76, Mariska+78, Sandlin+86, Curdt+01, Culhane+07}
Observations of the Zeeman effect in the C~{\sc iv} 1548\,\AA\ line
have in fact been used to measure the magnetic field in the TR over active regions 
\citep[ARs;][]{Tandberg+81, Henze+82, Hagyard+83}. %, Manso+19}.
Linear polarization in the S~{\sc i} 1437\,\AA\ line, most likely due to impact by 
beamed electrons, provided important insights into how flares work and 
the dynamics of these high layers \citep{Henoux+83}.

The Mg~{\sc ii} UV doublet at $\sim$2800\,\AA\ (h and k lines) was observed 
using full Stokes polarimetry 
%in active regions, looking for Zeeman dianostics of the
%magentic field in the chromosphere, and 
in quiet regions near the solar limb in search of scattering polarization \citep{HenzeStenflo87}.
Here, we present a reanalysis of those observations.  
The motivation for doing so stems from recent theoretical findings about the formation of 
broadband polarization in the Mg~{\sc ii} UV doublet.
Based on a thorough theoretical and modeling analysis,
\cite{delPino+16} predicted that a very broad spectral region $\sim$30\,\AA\ 
around the h and k lines should be highly linearly polarized and, surprisingly, 
{modulated by the chromospheric magnetic field via magneto-optical (M-O) effects}. 
We also analyze observations in ARs that appear to be dominated by the 
Zeeman effect.
\cite{delPino+16} and \cite{Alsina+16} have found that 
despite the complexity of the spectroscopic and transfer 
mechanisms involved in their formation, the circular polarization in the core
of the lines satisfies the magnetograph formula.
These findings have potentially important observational consequences. 
The former opens the interesting possibility of broadband filter polarimetry in the UV;
the latter justifies a simplified approach to the magnetometry in these 
highly complex layers of the Sun.

\section{Observations}

The observations were made with SMM/UVSP in 1980 March-April (Table 1), and 1984 October.
UVSP consisted of a Gregorian telescope, a spectrograph (Ebert-Fastie configuration),
and a rotating MgF$_2$ waveplate that could be inserted in the light path,
acting as the polarization modulator, while the grating itself served as the 
polarization analyzer. 

The detector was a single photocathode.
Spectral scans were made by rotating the grating of the spectrometer.
For each given wavelength and spatial position, a full 
polarimetric modulation cycle (consisting of 16 steps) was performed before
changing to a different wavelength or position.

A detailed description of the demodulation scheme and analysis of the observations
is given in the Appendix. 
Most of the analysis follows standard techniques
\citep[e.g.,][]{Woodgate+80, Tandberg+81, Henoux+83, HenzeStenflo87}; 
we present it here for completeness and to justify the particularities of our approach.
We improve upon previous analyses on three different aspects: recalibration of the 
linear polarization, analysis of the circular polarization, and a careful,
direct treatment of the error budget necessary for the statistical interpretation 
of the results.

\subsection{Limb and Disk-center Observations}

The observations were carried out at different wavelengths (up to 10, spanning 15\,\AA\ 
through the Mg~{\sc ii} h\,\&\,k doublet), with an entrance slit of 
1\arcsec$\times$180\arcsec, 
parallel to the solar limb, and an exit slit 0.02\,\AA\ wide. 
At each wavelength, the sun was rastered perpendicularly to the slit 
seven times at 10\arcsec % intervals---in the limb, radially outwards---,
before moving to a new wavelength. 
A complete modulation cycle lasted 40~s ($16\times 2.5$~s), 
and the full spatial raster at a given wavelength $\sim 5$~minutes.

The observational strategy was optimal for detecting linear (scattering) polarization at 
the 10 selected wavelength positions in the wings (two blueward of the k-line, 
three redward of the h-line) and between the lines (five). 
With the modulation scheme of UVSP, linear polarization (Stokes-$Q$ and $U$) 
appears at higher frequencies than circular polarization. 
Slow monotonic intensity drifts have more power at lower frequencies and 
may affect severely circular polarization signals,
while linear polarization, at higher frequencies, is more protected
from intensity-to-polarization cross-talk. 
In the longer exposures used for scattering at the limb,  
circular polarization was often vulnerable to such effects and 
those data rejected.
A more detailed discussion can be found in the Appendix.
The remaining reliable circular polarization observations (not shown in Fig.~\ref{fig:limb})
were compatible with null signal, as expected.

Observations at disk center were taken with the same instrumental configuration 
and observational strategy as for the limb observations, which allowed us 
to calibrate for spurious instrumental polarization \citep{Manso+17}--- 
a common strategy for inflight calibration even in modern spectropolarimeters 
\citep[e.g., ][]{Giono+17}. 
Averaging over all the slit positions and wavelengths, we find negligible amounts
of circular polarization and Stokes $U$, but a
residual signal $Q/I\sim 0.005$ (see Appendix). %Figure~\ref{fig01}, dotted lines).
All the observations reported here have been recalibrated to this $Q/I$ ``zero level''.
This step, not taken by \cite{HenzeStenflo87}, 
has important consequences for our confirmation of their tentative detection.

A scan along the whole range lasted up to $\sim$50\,min, far longer than
the typical dynamic timescales in the chromosphere and TR, thus 
precluding the reconstruction of resolved profiles.
Since the wavelength positions are reported in the logbooks \citep{Henze93} 
with very low accuracy, we adopted instead the values of \cite{HenzeStenflo87}.

\subsection{Active Region Observations}

A different strategy was followed to observe 
two ARs at two different positions on the solar disk, at two different times (Table~\ref{tab:log}).
The aim here was to scan the core and nearby wings where the Zeeman polarization peaks appear. 
The profile of the Mg~{\sc ii} k-line was scanned with a fixed spectral resolution of 
$\Delta\lambda\sim 100$\,m\AA, at a fixed spatial position. 
The entrance slit was 3\arcsec$\times$3\arcsec.
This was done twice, consecutively.

The integration time per wavelength was much shorter:
a modulation cycle lasted 16~s (16$\times$1~s). 
This was enough to significantly reduce the number of observations severely
affected by intensity drifts and   
also made it possible to reconstruct the line profile reliably (see Sect.~\ref{sect:4} below).
The total duration of the whole observation was $\sim 2\times 12$\,minutes.
Table~\ref{tab:log} summarizes the log record.

The demodulation and error analyses were otherwise performed as indicated in the Appendix.

\begin{table}
\centering
\caption{AR observation log and inferred magnetic field\label{tab:log}}
\begin{tabular}{cccc|rr}
\hline\hline
Exp.\tablenotemark{a} & AR id & R\tablenotemark{b} & Date   &  \multicolumn{2}{|c}{B\tablenotemark{c} ($\sigma$)} \\
 & NOAA &  &  & \multicolumn{2}{|c}{(G)} \\
\hline
00948 &  2363 & 0.5  &31.03.80   &  120 (200) & 69 (200) \\
01042 &       & 0.77 &03.04.80   &  331 (200) & 308 (200) \\
01182 &  2372 & 0.36 &07.04.80   &  10  (100) & 121 (100) \\
01237 &       & 0.08 &09.04.80   &  516 (200) & 158 (100) \\
\hline\hline 
\end{tabular}
\tablenotetext{a}{UVSP experiment number}
\tablenotetext{b}{Radius vector from disk center ($R=0$) to limb ($R=1$).}
\tablenotetext{c}{Magnetic field (in G) and $1\sigma$ uncertainty (between parenthesis) 
for each two consecutive modulations in a given experiment.
}
\end{table}

\section{Linear polarization in the wings: scattering at the limb}
%(line wings)

\begin{figure}
\centering
\includegraphics[width=9cm]{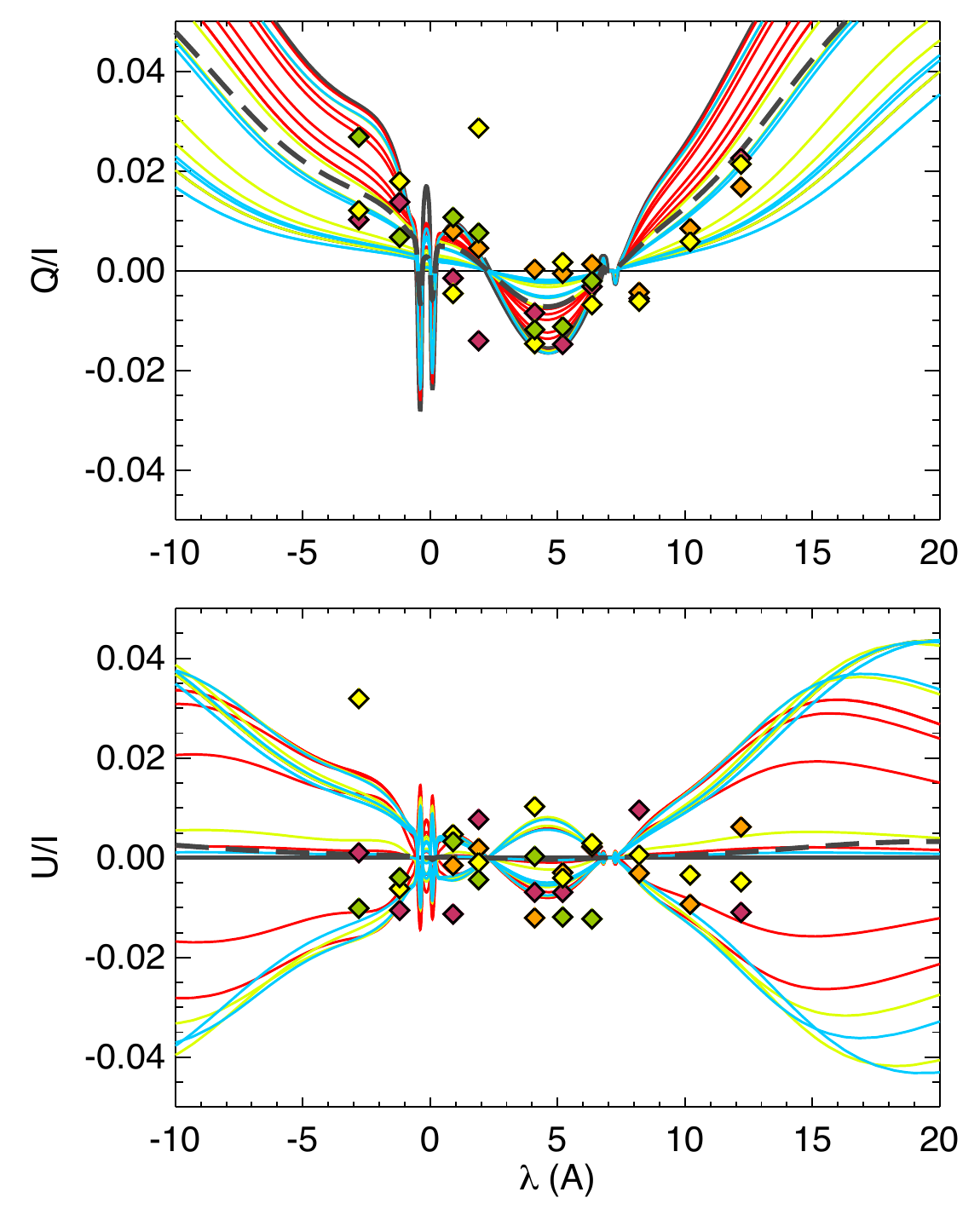}
\caption{\label{fig:limb}
Polarization observed at the limb ($\mu=0.15$). 
The symbols in four different colors correspond to four
sets of observations performed on the 9th (2), 1984 October 19th and 20th.
The scanning of a given ``profile'' required between 35 and 50 minutes. 
Solid lines illustrate the range of variability of the polarization pattern 
for different magnetic field configurations 
(red: $B=20$~G, inclined $\theta_B=60^\circ$ with respect to the local vertical; 
yellow: $B=50$~G, $\theta_B=60^\circ$; 
blue: $B=50$~G, $\theta_B=90^\circ$;
several different azimuths for each $B$ and $\theta_B$ are shown).
Dashed line: random azimuth field ($B=50$~G, $\theta_B=45^\circ$).  
The thick dark line: polarization corresponding to the 
nonmagnetic, pure scattering case.
The theoretical calculations are described in 
\cite{delPino+16, Manso+17}.
}
\end{figure}

\begin{figure}
\centering
\includegraphics[width=9cm]{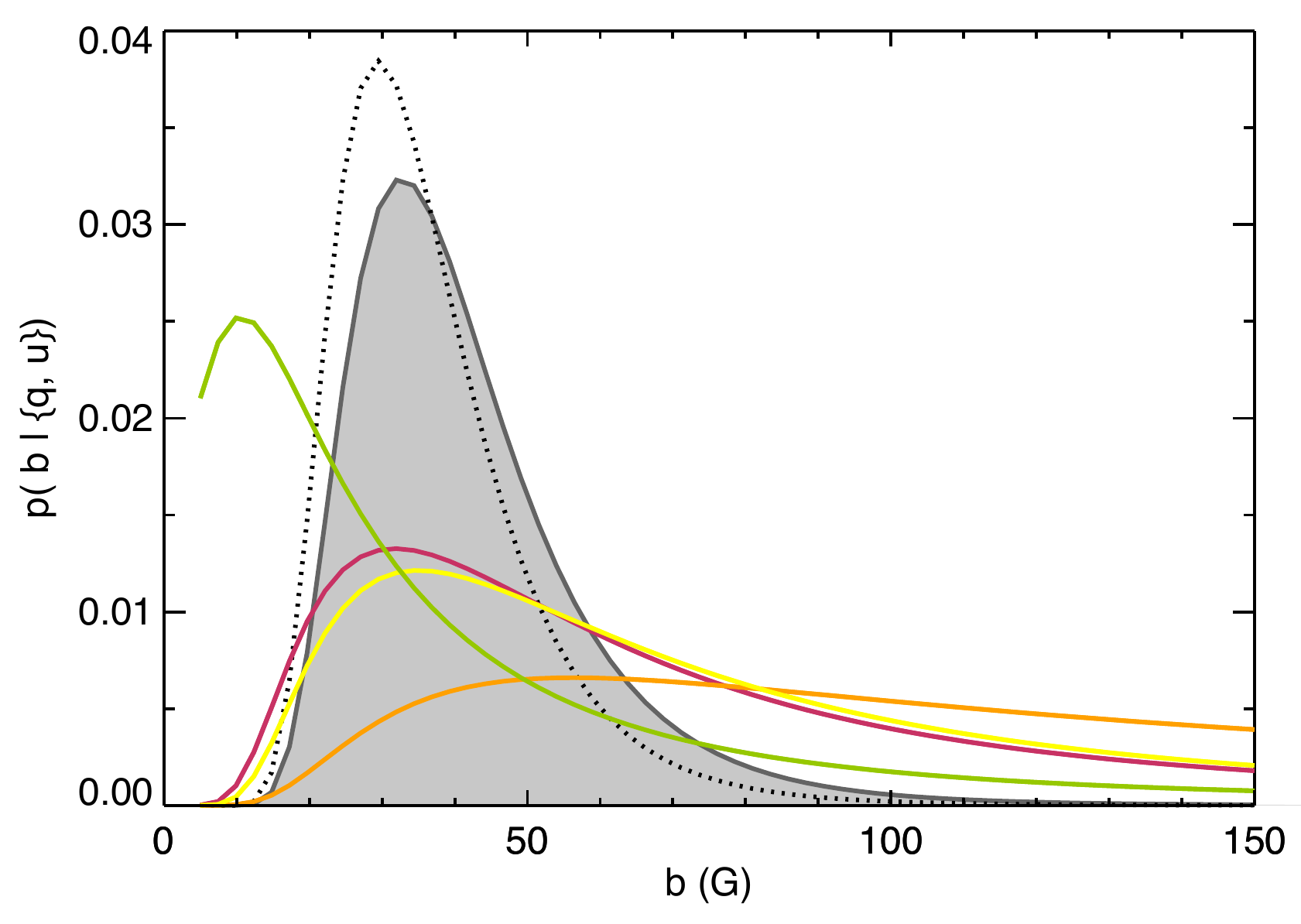}
\caption{\label{fig:post2}
Posterior of the hyper-parameter $b$ characterizing the Maxwellian distribution 
of fields adopted for the individual observations. 
Each color corresponds to a data set in Fig.~1.
A joint analysis assumed the same Maxwellian distribution for all the observations
and a uniform prior (shaded area) or Jeffreys prior (dotted) for $b$.
A Maxwellian distribution with $b\approx 30$\,G is preferred as MAP, which has
an average field of $B\approx50$\,G.
}
\end{figure}

Observations close to the limb were carried out motivated by theoretical 
calculations predicting very high polarization levels due to scattering 
in the wings of the Mg~{\sc ii} doublet, and a significative negative 
polarization lobe between the h and k lines \citep{Auer80}. 
The results were deemed only partially successful; 
a polarization increase toward the far wings was indeed clearly detected but 
the negative lobe could not be confidently confirmed \citep{HenzeStenflo87}.

Our reanalysis definitely confirms the presence of the negative 
polarization lobe (Figure~\ref{fig:limb}).
By contrast, we find a lower degree of polarization in the 
wings---slightly below, in fact, than what would be expected in the absence of 
magnetic fields (pure scattering case). 
However, this is precisely consistent with the recent theoretical finding on the magnetic
modulation of the broadband polarization pattern around the Mg~{\sc ii} UV doublet 
due to M-O effects in the far wings  
\citep{delPino+16, Manso+17}.
Figure~\ref{fig:limb} shows that the observed fractional polarizations $q=Q/I$ and $u=U/I$
(previously unreported), fall nicely within the range of values allowed in a magnetized
atmosphere: $|q|$ below the the pure scattering case, $|u|\geqslant 0$.

The spectropolarimetric pattern in a spectral region $\pm$20\,\AA\ around
the Mg~{\sc ii} UV doublet has been studied by \cite{delPino+16} 
in great generality.
Their modeling includes all the relevant spectroscopic and polarigenic
mechanisms involved in the formation of the observed patterns---in particular, 
partial frequency redistribution (PRD), quantum coherence among the levels of the 
upper term ${}^2\!P$, and, crucially, M-O effects. 
%\textbf{\emph{(I'd add the paper on the heuristic derivation of the branching ratios.)}}
The calculations presented here were made for a semiempirical model atmosphere 
of the quiet sun \citep[model C of][]{Fontenla+93}, for a grid of 
magnetic fields up to 1500\,G, constant with height, different inclinations
and azimuths with respect to the line of sight (LOS). 
Further details on the numerical methods, the code, and a detailed discussion on 
the physical processes involved can be found in \cite{delPino+16,Casini+17}.

It is not possible to uniquely determine the full magnetic field vector, ${\boldsymbol B}$,
from just two observables, $q$ and $u$.
One may, however, infer information on a single parameter---say, the magnetic
field strength, $B$---probabilistically with high confidence, 
by considering all the possibilities compatible with the observations (within
uncertainties) and their probabilities, and marginalizing over all other 
less interesting (or uncertain) ``noise'' parameters. 
This Bayesian approach is the most natural and optimal way to extract 
the magnetic information from the data.
%\textbf{\emph{(this 
%last statement comes a bit too late, as you have already talked of priors, 
%marginalization, etc...)}}

The observations at the limb were performed with very low 
%\textbf{little (? or do you mean that itwas either very high or very low?)} 
(or none) spatial and temporal resolution,  
and they do not resolve the characteristic spatio-temporal dynamical evolution of the 
magnetic field on the chromosphere. 
So they have little information on the magnetic field vector configuration.
The simplest model to describe this scenario is a macroscopically random azimuth 
magnetic field (in the local reference frame). 
Indeed, we expect the azimuth 
of the magnetic field to change much more widely during an observation 
than the inclination and polarity of the field.
The synthetic averaged polarization profiles $\bar{q}$ and $\bar{u}$  
depend on the magnetic field strength $B$ and inclination $\theta_B$, 
but we will marginalize over $\theta_B$ at the end of the 
calculation.\footnote{Alternatively, 
including explicitly both the azimuth $\chi_B$ and inclination $\theta_B$
in the model [$q(B, \theta_B, \chi_B)$, $u(B, \theta_B, \chi_B)$], and then 
marginalizing over both parameters, would correspond to the case in which 
the magnetic field is (mostly) resolved in each individual observation 
and then combined {\em a posteriori} to statistically constrain the field.
This is different from our case. 
Our observations do not resolve the magnetic geometry; 
we keep $\theta_B$ as the minimal parameter for marginalization
to describe some degree of resolution.}
This is a very simplified minimal model of the observations, yet
it will allow a quantitative estimate of the magnetic field in the chromosphere.

For simplicity, we compare observations directly with the results of the numerical 
synthesis. 
Considering a finite spectral resolution does not noticeably affect the results
due to the sparse sampling in wavelength and because the profiles are 
spectrally smooth in the wings. 
The relatively large uncertainties on the observations (wavelength position)
are already compatible with neglecting these effects. 
This, however, would not be the case for high resolution spectral and spatial observations.

For a given set of observations $\{q_\ell, u_\ell\}$, $\ell=1,\ldots, N_\ell$,  
assuming statistical independence among the $N_\ell$ wavelengths and Gaussian noise, 
we build a likelihood function
${\cal L}(B, \theta_B)\equiv p(\{q_\ell, u_\ell\}|B,\theta_B)={\rm e}^{-\chi^2}/Z$,
with merit function
\begin{equation}
\chi^2 = \sum_\ell \frac{[q_\ell-\bar{q}_\ell(B,\theta_B)]^2 
+ [u_\ell-\bar{u}_\ell(B, \theta_B)]^2}{2\sigma_\ell^2}\;,
\end{equation}
and normalization constant %({\em Zustandssumme}) 
$Z=(2\pi)^{N_\ell/2}\prod_\ell \sigma_\ell$, 
%\textbf{\emph{(are you sure this is $Z^{-1}$ and not directly $Z$?)}}
where $\bar{q}_\ell(B, \theta_B)$ and $\bar{u}_\ell(B, \theta_B)$ are calculated numerically, 
and $\sigma_\ell^2 \equiv\sigma^2(q_\ell)=\sigma^2(u_\ell)\approx \sigma^2(Q)/I = \sigma^2(U)/I$,
is calculated as described in the Appendix.
Assuming minimal {\em a priori} information on the magnetic field inclination,
(e.g., isotropy),
%; hence, uniformity on $\cos(\theta_B)$), 
we marginalize over
$\theta_B$ to obtain ${\cal L}(B)=\int {\cal L}(B, \theta_B)\,d\cos\theta_B$.
The maximum-likelihood estimates for the four sets of observations
are  20, 48, 50, and 70\,G.

We may improve this estimate using the Bayes theorem 
and a simple hierarchical scheme assuming that the measured value $B$
stems from a population with a Maxwellian distribution, $B\sim {\cal M}(b)$.
The parameter $b$ 
($=\sqrt{\pi/8}$ times the mean of the distribution), 
will act as a hyper-parameter.
The posterior for $b$ is then $p(b|\{q_\ell, u_\ell\})\propto \int {\cal L}(B)p(B|b)p(b)\,dB$;
it is shown in Fig.~\ref{fig:post2}, calculated for each of the four sets of observations.

Finally, we may pool all the observations assuming a common underlying distribution
(hence, the same $b$) for all of them.
We find a maximum {\em a posteriori} (MAP) for $b$ 
of 29\,G using an uninformative improper Jeffreys prior $p(b)=1/b$ \citep[e.g.,][]{Jaynes03},
or 32\,G for a uniform prior (see Fig.~\ref{fig:post2}).
This is corresponds to an average field $B\approx 46$\,G.

\section{Circular polarization in the line core: Zeeman effect in active regions \label{sect:4}}
%(line core)

\begin{figure}
\centering
\includegraphics[width=8cm]{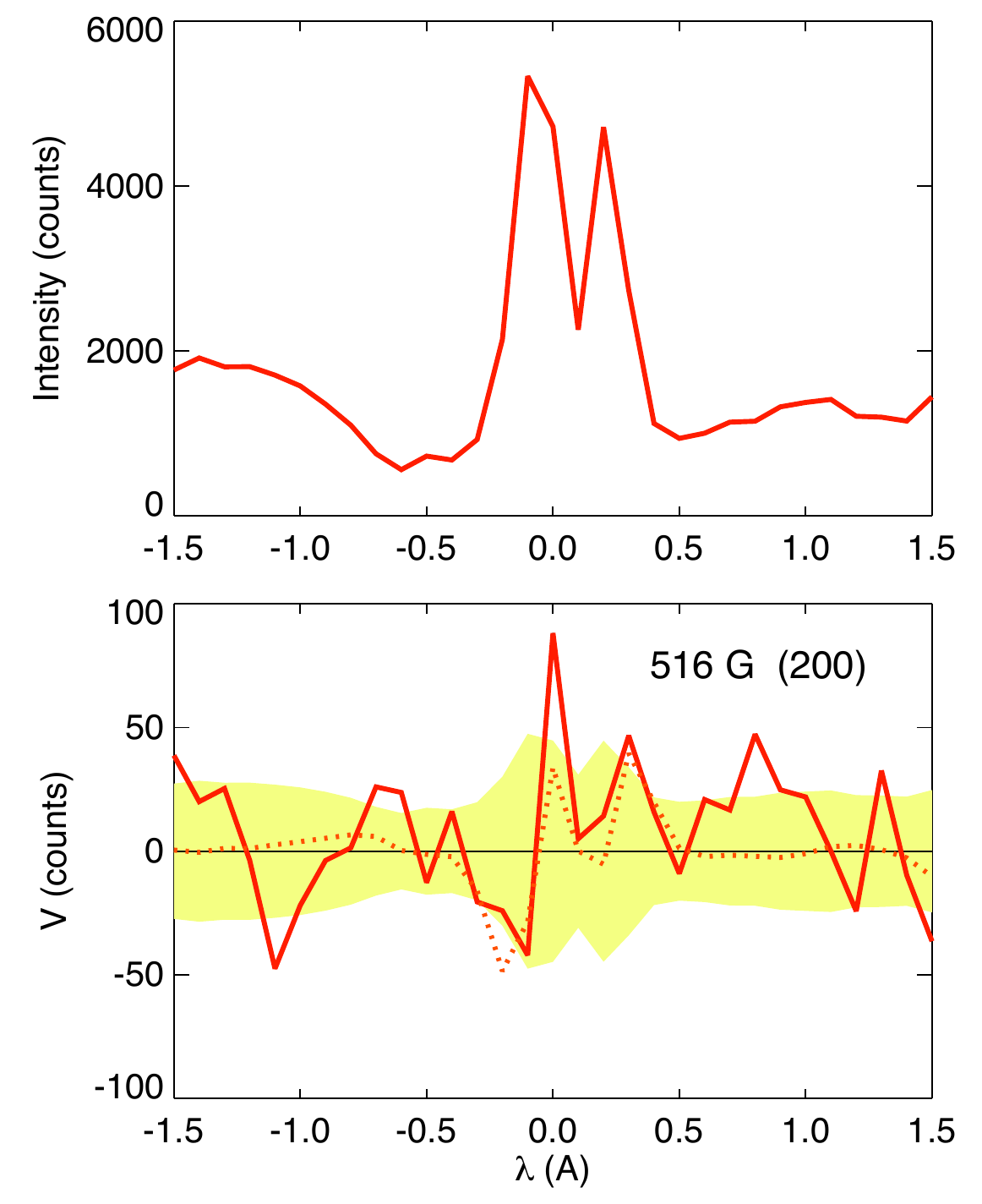}
\caption{\label{fig:AR}
Intensity (upper panel) and circular polarization (lower panel) 
of the Mg~{\sc ii} k-line observed 
in active region NOAA 2372 at the center of the solar disk 
(see entry for Exp. 01237 in Table~1).
The yellow shaded area in the lower panel represents the $\sigma(V)$ level.
The dotted line shows the fitting of the spectral derivative of the intensity 
to $V$.
A magnetic field of 516~G ($\sigma(B)=200$~G) is inferred according to 
the magnetograph formula.
}
\end{figure}

A simple, direct estimate of the magnetic field along the LOS, $B_\parallel$, 
is obtained from the observed intensity and circular polarization in a line profile,
using the simple relation \citep{LandiLandi72}
\begin{equation}\label{eq:magnetograph}
V_\lambda=-C B_\parallel \frac{dI_\lambda}{d\lambda}\;,
\end{equation}
where $C=\lambda_0^2\; g_{\rm eff}\, e /(4\pi m c^2)\approx 4.6686\times 10^{-10} 
(\lambda_0/{\text \AA})^2\;g_{\rm eff}$~m\AA\ G$^{-1}$
(here symbols have their conventional meaning, and $g_{\rm eff}$ is the effective Land\'e factor).  
In our case, $\lambda_0=2795$\,\AA, and $g_{\rm eff}=7/6$.
The \emph{magnetograph formula} (\ref{eq:magnetograph}) is valid in the weak 
magnetic field limit, but it is a good approximation in the core of the Mg~{\sc ii} k-line
for the magnetic field strengths expected in the upper solar chromosphere \citep{delPino+16, Alsina+16}. 
%

%We build the likelihood function for model (\ref{eq:magnetograph}) 
%approximating observational noise as Gaussian,  
%${\cal L}\propto\exp(-\chi^2)$, with the merit function\footnote{One should consider that both observations, $V$,
%and ``model'' ($-C B_\parallel I'$) have observational 
%uncertainties---even correlations may exist \citep[see][]{AsensioManso11}.
%This may be safely neglected for $(CB_{\parallel}/2\Delta)^2\sigma^2(I)\ll\sigma^2(V)$,
%i.e., $B_{\parallel}\ll 2\Delta/|p\sin\delta|\approx 20$~kG ($\Delta=3$-10 m\AA).}, 
%\begin{equation}
%\chi^2 = \sum_l \frac{[V_l+C \, B_\parallel\,I_l']^2}{\sigma^2(V_l)},
%\end{equation}
%where the sumation on $l$ extends over all the sampled wavelengths
%and the derivative $I'=dI/d\lambda$ is calculated numerically.
%We maximize ${\cal L}$ minimizing $\chi^2$ and find the
Applying a maximum-likelihood argument to Eq.~(\ref{eq:magnetograph}),
we find an estimate for the longitudinal magnetic field 
and its uncertainty \citep{marian12},
\begin{subequations}\label{eq:MLB}
\begin{gather}
B_\parallel = -\frac{1}{C}\frac{\sum_l {V_l I_l'}/{\sigma^2(V_l)}}{\sum_l {(I_l')^2}/{\sigma^2(V_l)}}\;, \\
\sigma^2(B) = \frac{1}{C^2}\frac{1}{\sum_l {(I_l')^2}/{\sigma^2(V_l)}}\;,
\end{gather}
\end{subequations}
where the derivative $I'=dI/d\lambda$ is calculated numerically.
%\citep[e.g.,][Section 9.6]{LandiLandolfi04}.
Noise at each wavelength is explicitly taken into account in Eqs.~(\ref{eq:MLB}),
which is particularly important for emission lines like this.

The values of the magnetic field inferred by applying Eqs.~(\ref{eq:MLB}) to 
the eight UVSP observations of two ARs are tabulated in Table~\ref{tab:log}.
In all but one case, there is no magnetic field detection above the 
the 2$\sigma$ level.
The observation of NOAA\,2372 at disk center however, reaches 516\,G at a 2.5\,$\sigma$ level
(Fig.~\ref{fig:AR}).
Interestingly, NOAA\,2372 was an exceptionally energetic AR during Solar Cycle 21, 
consisting of two large opposite polarity sunspots with a smaller bipole between them.
It was reported flaring more than 50 times during its transit over the disk
\citep{Sawyer82, Machado83, Ambastha88},
and it showed an unusual evolution of the magnetic field in the photosphere
\citep{Krall82, Rabin84}.
Both factors---the favorable disk center geometry and the exceptionally strong activity---make it plausible that a magnetic field as high as 500\,G 
was present at some point in the high chromospheric layers probed by 
the core of the Mg~{\sc ii} k line.

\section{Discussion}

Our reanalysis of the linear polarization in the wings of the Mg~{\sc ii} h and k lines
observed close to the solar limb confirms the presence of large (positive) signal in
the outer wings and a negative polarization between the two lines.
This conclusion differs from a previous analysis by \cite{HenzeStenflo87}, both quantitatively
(they found an even larger polarization degree in the wings) and qualitatively (they found
their results compatible with virtually no polarization between the two lines).
Such discrepancy is resolved by our recalibration of the linear polarization.
This reveals clearly the presence of the negative lobe, while at the same time it shows 
linear polarization in the wings that are below the theoretical zero-field level. 
This last result is explained as a combination of M-O effects in the wings \citep{delPino+16, Manso+17}
and a lack of adequate spatio-temporal resolution.
%\textbf{(spectral? spatial? temporal? all of the above?)}. 
A simple statistical analysis suggests a magnetic field $B\approx 50$\,G 
in the region between $\sim$500-700\,km of height where the observed 
line wings form (del Pino Alem\'an et al., submitted).

The analysis of the circular polarization produced by the Zeeman effect 
\citep{delPino+16, Alsina+16}
shows no significant detection above $\sim 300$\,G in observations above ARs, except in
one instance of particular interest: in NOAA\,2372 we detected a field of 516\,G at the 
2.5\,$\sigma$ level.

These results support the most recent theoretical understanding of the formation of the 
Mg~{\sc ii} UV doublet, the validity of the current numerical modeling, and can be of 
great interest for the future analysis of spectropolarimetric observations in this 
important magnetic diagnostic of the solar chromosphere and TR.

\acknowledgements 

We thank Robert M. Candey (NASA), Joseph B. Gurman (NASA), and Juan Fontenla, for
stewarding legacy data and codes through the times and kindly helping us to locate them.
We also thank the referee for comments and suggestions that helped us 
present more clearly the analysis of this unconventional, venerable data set.
TdPA acknowledges funding from the European Research Council (ERC) under
the European Union's Horizon 2020 research and innovation programme 
(grant agreement no. 742265).
This material is based upon work supported by the National Center for Atmospheric Research, which is a major facility sponsored by the National Science Foundation under Cooperative Agreement No. 1852977.

\appendix

\section{Data Analysis}

The polarimetric modulation of the UVSP consisted of rotating the 
MgF$_2$ waveplate 16 times at
22.5$^\circ$ intervals. The (polarizing) grating of the spectrograph 
acted as the polarization analyzer.
The observed modulated intensities can then be modeled as \citep{Calvert+79}
\begin{equation}\label{eq01}
S_j = I + p[a+b\cos(4\theta_j)]Q 
-p \,b \sin (4 \theta_j) U + p \sin\delta\sin(2\theta_j)V\;,
\qquad j=0, \dots,N-1\;, 
\end{equation}
where 
$\theta_j=2\pi j/N$ ($N=16$ in our case) 
$\delta$ is the retardance of the waveplate,
$p$ is the polarization efficiency of the analyzer, and 
$a=(1+\cos\delta)/2$, $b=(1-\cos\delta)/2$ 
\citep[at 2800\,\AA, $p=0.72$, $\delta=131^\circ$,
from laboratory measurements before launch;][]{HenzeStenflo87}.
The Stokes parameters $I, Q, U, V$ of the incoming 
light are defined in the usual way \citep[e.g., ][]{LandiLandolfi04}.
The positive direction for $Q$ was along the solar North-South line, 
which in our case was parallel to the limb. %%%%%%  {\bf !!!!!!}

In practice, there may be intensity drifts during the observation due
to the finite duration of the modulation scheme ($\sim 5$\,minutes), which
are not taken into account by Eq.~(\ref{eq01}).
We corrected for them by fitting a linear trend to the $S_j$ series 
and substracting it, preserving the average. 
This produces typically small corrections with respect to \cite{HenzeStenflo87}.

\begin{figure*}
\centering
%\sidecaption
\includegraphics[height=15cm]{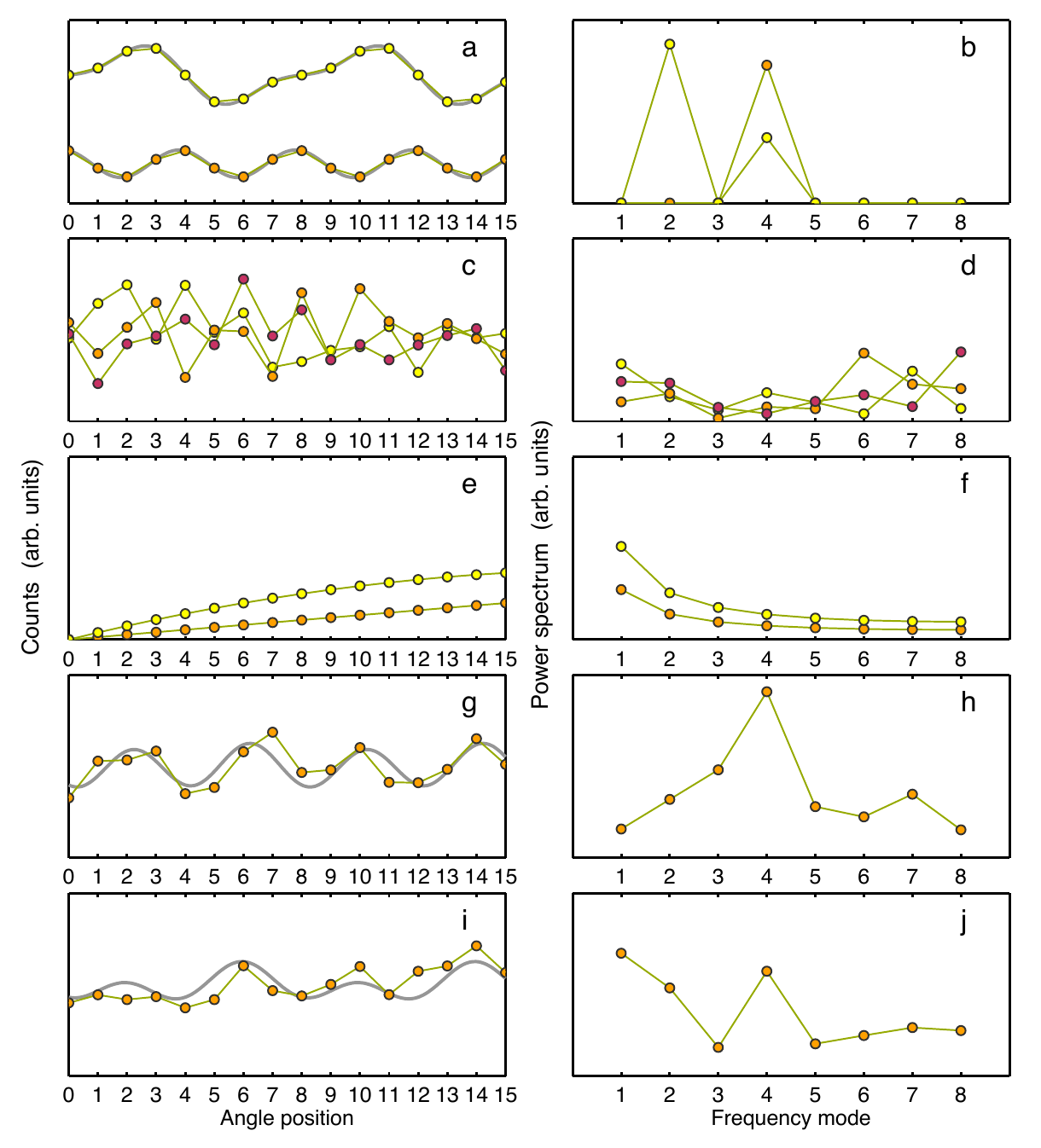}
\caption{\label{fig00}
Modulated signals $S_j$ in time space (left column), and corresponding power 
spectrum $|s_k|^2$ in Fourier space (right column);
the redundant${}^2$ Fourier modes 9-15 and the (much stronger) 0-th mode 
are not shown for clarity.
a, b) Ideally, only modes 2 and 4, corresponding to circular and linear polarization, 
respectively, would be present in the signal. 
Here, two ideal cases are shown as illustration: 
a linearly polarized signal (orange) and an elliptically polarized  
(linear and circular components; yellow) signal.
%\textbf{\emph{(the verb is missing...)}}.
c, d) Random (white) noise introduces spurious power in all the modes.
e, f) Intensity drifts---e.g., linear (orange), parabolic (yellow)---during the 
observation and other ``slow'' trend artifacts introduce more power at lower frequencies.
g, h) Actual, observed noisy linearly polarized signal. 
Gray line: best fit reconstructed signal.
i, j) Observed noisy signal with significant drift. 
Notice that the linear polarization mode ($k=4$)
is above the noise, but the circular polarization mode ($k=2$) is below the  
mode $k=1$, suggesting that it is also likely affected by drift.
Panels g)-j) show raw, uncorrected (for intensity drift) data. 
}
\end{figure*}

The Stokes parameters at a given wavelength and spatial position are 
recoverd from the $S_j$ through least-squares fitting or, 
equivalently, by discrete Fourier analysis (Figure~\ref{fig00}),
\begin{equation}\label{eq02}
S_j=\sum_{k=0}^{N-1}s_k \exp\biggl(2\pi {\rm i}\,\frac{jk}{N}\biggr)\;, 
\qquad\qquad
s_k=\frac{1}{N}\sum_{i=0}^{N-1}S_j \exp\biggl(2\pi {\rm i}\,\frac{jk}{N}\biggr)\;.
\end{equation}
The $s_k$ coefficients can be efficiently calculated by the Fast Fourier Transform
algorithm \cite[e.g.,][]{Press+92}.
Then\footnote{Because the observed intensities are real 
($S_j=S_j^*$, where ${}^*$ indicates complex
conjugation), in our case $s_k=s_{16-k}^*$. Hence, 
$S_j=s_0+2\sum_{k=1}^7[\Re(s_k)\cos(\pi jk/8)-\Im(s_k)\sin(\pi jk/8)]
+s_8$.} 
%\textbf{\emph{(huh? what is $n$?)}}}  
(cf.\ Eqs.~(\ref{eq01}), (\ref{eq02})),
\begin{equation}
I = s_0 - 2\, \frac{a}{b}\, \Re(s_4)\;, \qquad
V = -\frac{2}{p\sin\delta}\, \Im(s_2)\;, \qquad
Q = \frac{2}{pb}\, \Re(s_4)\;, \qquad
U = \frac{2}{pb}\, \Im(s_4)\;. 
\end{equation}
Note that only modes $s_{2, 4}$ (and the {\em average} $s_0$) contribute to the signal.
%\textbf{\emph{(in the figure you called $s_0$ ``spike''... I'd use a consistent 
%terminology)}}

The uncertainty of the modulated intensities at each wavelength is assumed to be
Poissonian (i.e., $\sigma^2(S_j)\equiv\sigma^2(S)=\bar{S}$, where $\bar{S}$ is the average of the $S_j$.
Then, by error propagation, 
$\sigma^2(s_0)=\sigma^2(S)/N$, $\sigma^2(\Re(s_k))=\sigma^2(\Im(s_k))=\sigma^2(S)/2N$
($0<k<8$), and therefore, 
%\textbf{\emph{(what is $\sigma^2(S)$ in all these formulas? shouldn't be simply $\bar S$?)}}
\begin{equation}
\sigma^2(I) \approx \frac{\sigma^2(S)}{N} ,  \qquad
\sigma^2(V) = \frac{2}{(p\sin\delta)^2}\frac{\sigma^2(S)}{N}, \qquad
\sigma^2(Q) = 
\sigma^2(U) = \frac{2}{(pb)^2}\frac{\sigma^2(S)}{N}.  
\end{equation}

In some observations, a few pixels were lost due to telemetry glitches
and only 14 or 15 $S_j$ values were recovered in a given series; 
in those cases, a least-squares fitting was applied.

Random (white) noise introduces spurious power in all the modes---not just $k=2$ and 4. 
Additionally, ``slow'' intensity drifts during the observation introduce 
more power at lower fequencies (Figure~\ref{fig00}).
Typically, we find that the noise from an average of modes $k=3$ and 5-8, 
is consistent with the noise estimate given above assuming photon (Poisson) noise. 
In some cases, however, the power in mode $k=1$ is clearly well above the average 
noise. 
We consider this as an indication of systematic artifacts in that observation,
which likely affect also the mode $k=2$; the circular polarization measurement
in those cases is discarded, while the mode $k=4$ (linear polarization) is 
typically protected from these drifts (see panels i-j in Figure~\ref{fig00}).

\begin{figure}
\centering
\includegraphics[width=9cm]{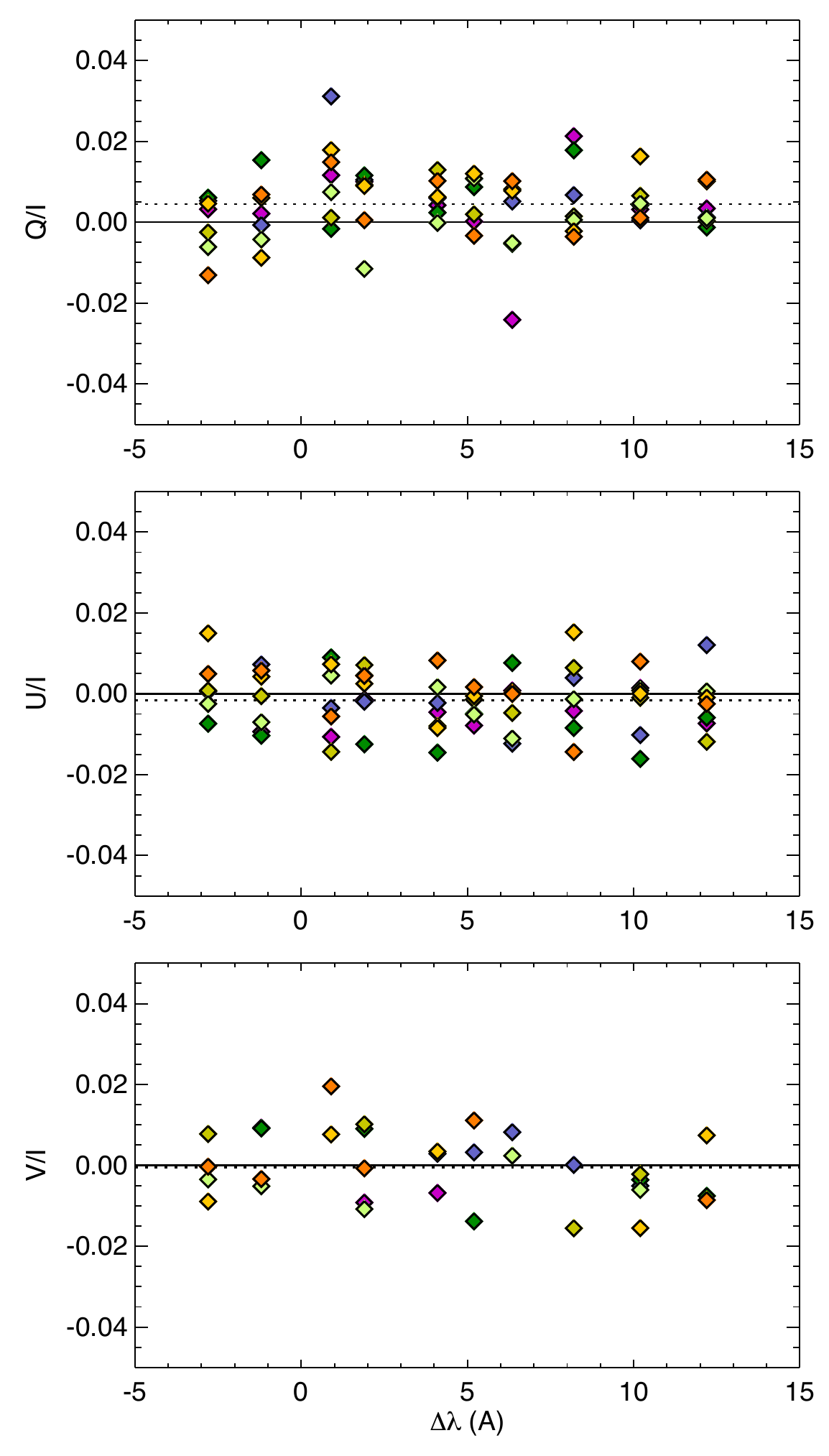}
\caption{\label{fig01}
Polarization at disk center on October 30th, 1984 (01:36-02:24). 
Each color corresponds to a given spatial position (up to seven at a given wavelength).
%All the observations at a given wavelength were taken simultaneously,
Between succesive wavelengths, $\sim 5$\,min elapsed (scan proceeded from blue to red).
Dotted lines show averages over all positions and wavelengths
	(wavelength scale centered at Mg~{\sc ii} k-line).
}
\end{figure}

Figure~\ref{fig01} shows the polarization observed at disk center 
with an observational setup identical to the limb observations of scattering
polarization.
We use the average at each Stokes parameter 
over all wavelengths and spatial positions as a calibration for the 
zero-polarization level---symmetry requires that a long enough integration 
over a wide area should show no polarization at any wavelength.
Only Stokes $Q$ needs to be significantly recalibrated.

\bibliography{ms2.bib}{}
\bibliographystyle{aasjournal}

%\bibliographystyle{apj}
%\bibliography{ms2.bib}

\end{document}